\documentclass[twocolumn,prl,aps,superscriptaddress,floatfix,
tightenlines,longbibliography]{revtex4-1}
\usepackage{epsfig,dsfont,amssymb,amsmath,amsthm,amsfonts,amsbsy,mathrsfs}
\usepackage{amsmath}
\usepackage{amssymb}
\usepackage{mathrsfs}
\usepackage{amsfonts}
\usepackage{epsfig}
\usepackage{bm}
\usepackage{times}
\usepackage{verbatim}
\usepackage[T1]{fontenc}
\usepackage[dvips]{color}
\usepackage[colorlinks,bookmarks=false,citecolor=blue,linkcolor=magenta,urlcolor=blue]{hyperref}
\usepackage{graphicx}
\begin{document}
\title{ High-Pressure NMR Enabled by Diamond Nitrogen-Vacancy Centers}

\author{Yan-Xing Shang}
\thanks{These authors contributed equally to this work.}
\affiliation{Institute of Physics, Chinese Academy of Sciences, Beijing 100190, China}
\affiliation{School of Physical Sciences, University of Chinese Academy of Sciences, Beijing 100049, China}

\author{Fang Hong}
\thanks{These authors contributed equally to this work.}
\affiliation{Institute of Physics, Chinese Academy of Sciences, Beijing 100190, China}
\affiliation{Songshan Lake Materials Laboratory, Dongguan, Guangdong 523808, China}

\author{Jian-Hong Dai}
\affiliation{Institute of Physics, Chinese Academy of Sciences, Beijing 100190, China}

\author{Ya-Nan Lu}
\author{Hui Yu}
\author{Yong-Hong Yu}
\affiliation{Institute of Physics, Chinese Academy of Sciences, Beijing 100190, China}
\affiliation{School of Physical Sciences, University of Chinese Academy of Sciences, Beijing 100049, China}

\author{Xiao-Hui Yu}
\email{yuxh@iphy.ac.cn}
\affiliation{Institute of Physics, Chinese Academy of Sciences, Beijing 100190, China}
\affiliation{Songshan Lake Materials Laboratory, Dongguan, Guangdong 523808, China}

\author{Xin-Yu Pan}
\email{xypan@aphy.iphy.ac.cn}
\affiliation{Institute of Physics, Chinese Academy of Sciences, Beijing 100190, China}
\affiliation{Songshan Lake Materials Laboratory, Dongguan, Guangdong 523808, China}
\affiliation{CAS Center of Excellence in Topological Quantum Computation, Beijing 100190, China}

\author{Gang-Qin Liu}
\email{gqliu@iphy.ac.cn}
\affiliation{Institute of Physics, Chinese Academy of Sciences, Beijing 100190, China}
\affiliation{Songshan Lake Materials Laboratory, Dongguan, Guangdong 523808, China}
\affiliation{CAS Center of Excellence in Topological Quantum Computation, Beijing 100190, China}


\begin{abstract}
The integration of NMR and high pressure technique brings unique opportunities to study electronic, structural and dynamical properties under extreme conditions. Despite a great degree of success has been achieved using coil-based schemes, the contradictory requirement on sample volume of these two techniques remains an outstanding challenge.
In this letter, we introduce diamond nitrogen-vacancy (NV) centers, as the source and probe of in-situ nuclear spin polarization, to address the sample volume issue.
We demonstrate hyperpolarization and coherent control of $^{14}$N nuclear spins under high pressures.
NMR spectra of a micro-diamond are measured up to 16.6 GPa, and unexpected pressure shift of the $^{14}$N nuclear quadrupole and hyperfine coupling terms are observed.
Our work contributes to quantum sensing enhanced spectrometry under extreme conditions.

\end{abstract}

\maketitle
The combination of  nuclear magnetic resonance (NMR) and high pressure techniques provides a powerful tool to explore the frontier of condensed matter physics, material science, geoscience and life science. On the one hand, pressure is a basic thermodynamical parameter and pressure tuning is widely used in natural sciences.
 For example, high-pressure conditions are indispensable for some material synthesis and properties optimization processes \cite{HPHTDiamond, LaH10_HP_PRL, LaH10_HP_Nature}.
High pressure brings exotic states like unconventional superconductivity and magnetic quantum critical point to condensed matter physics \cite{2020ReviewHPSC, MnP_SC}.
On the other hand, NMR is an important spectrum methods to reveal local electronic, chemical and structural information.
High-pressure NMR offers unique opportunities to study dynamical structure of hydrogen and hydrides \cite{HPNMR_for_H, H_NMR200GPa}, phase transitions of metal and superconductor \cite{HPNMR_review2018}, and protein folding process \cite{2017Review_HP_protein}, to name a few.

Despite impressive advances have been made, high-pressure NMR still faces a fundamental challenge, that is, the contradictory requirement on sample volume of these two techniques. Diamond anvil cells (DAC) is the widely used scheme to prepare and maintain high-pressures conditions in lab, but it provides only pico- to nano- liters sample volume \cite{HPNMR_review2018}. As an example, a DAC with culet of 500-$\mu$m diameter can sustain pressure of 30 GPa and provide a sample volume of several nanoliters,  and higher pressures corresponding to smaller sample volume.
For coil-based NMR measurements, due to the large thermal noise from the coil and weak gyromagnetic ratio of nuclear spins (thus low polarization rate), a relative large sample volume (> nanoliters) is usually required to build a reasonable signal-to-noise ratio (SNR) \cite{NMR_nl_1995}.
The direct integration of traditional NMR and high-pressure techniques requires special design of pick-up coils, which can fit the geometrical restrictions and sustain high pressures inside DAC. In this direction, a series of delicate coils have been designed to implement high-pressure NMR, including split-pair coils \cite{1987DAC_NMR}, gasket resonators \cite{1998DAC_NMR}, and micro-coils \cite{micro_coils}. Recently, Lenz lenses with near unit filling factors and enhanced sensitivity enable NMR at megabar pressure \cite{Lenz_lenses, H2O_NMR100GPa, H_NMR200GPa}.

\begin{figure}[b]
	\includegraphics[width=0.48\textwidth]{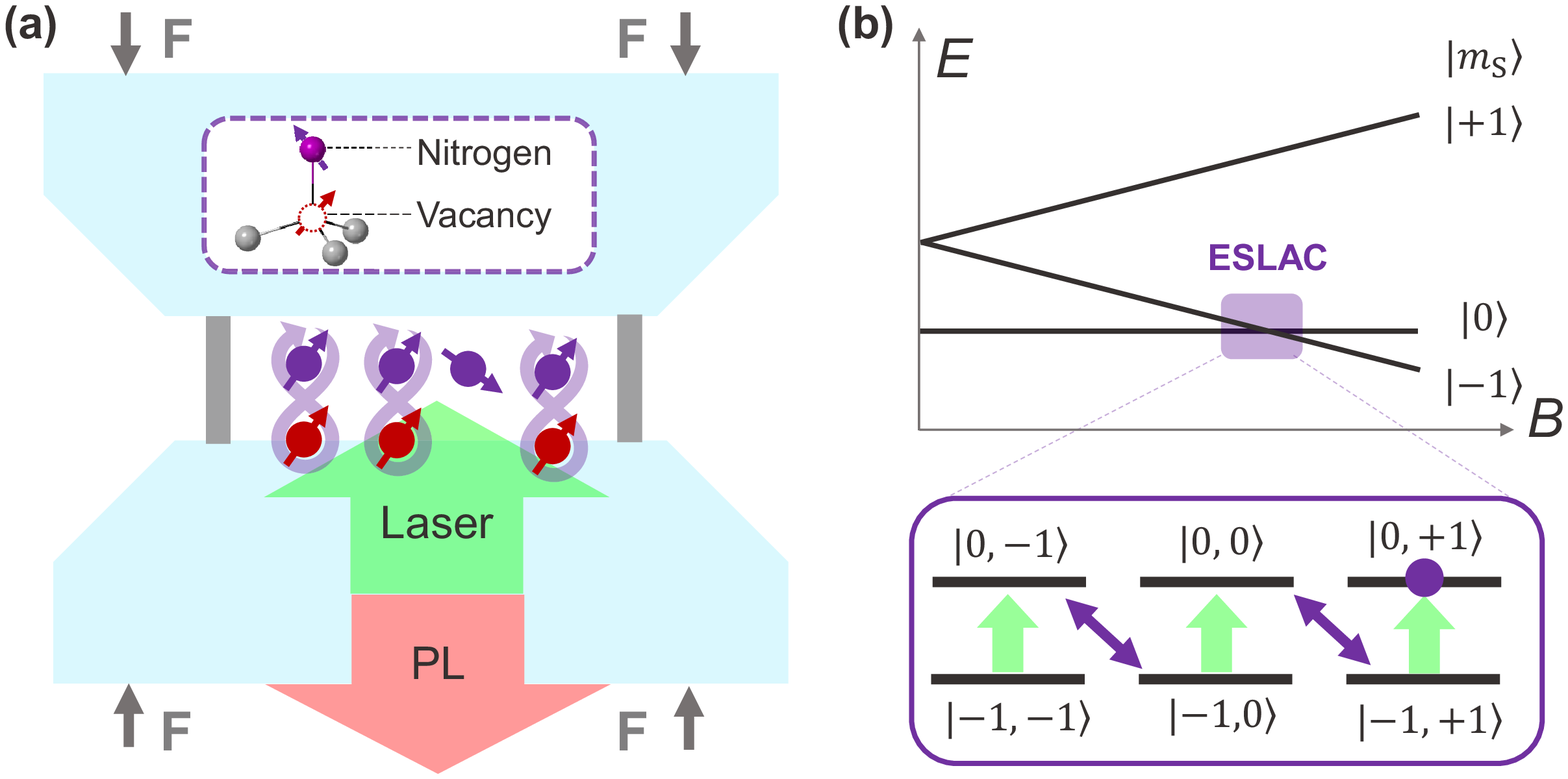}
	\caption{(color online). Optical hyperpolarization and readout of nuclear spins inside diamond anvil cells. (a) Nuclear spins (purple) are usually immune to optical excitation, but can be optically polarized (dynamic nuclear spin polarization, DNP) and read out through nearby nitrogen-vacancy (NV) centers,  provide a new strategy for  implementing NMR under high pressures. (b) An external magnetic field of about 500 Gauss brings NV center to its excited state level anti-crossing (ESLAC), where a ~$\mu$s laser pulse polarize both the NV electron spins and nearby nuclear spins. The same mechanism enables optical readout of the $^{14}$N nuclear spin states. }
	\label{fig1}
\end{figure}

Recent advances of quantum sensing via diamond nitrogen-vacancy (NV) center shine new light on the development of high-pressure NMR.
Hosted by the diamond lattice, NV centers present superb spin coherence and efficient optical interface, makes them ideal quantum sensors of magnetic fields under high pressures \cite{PRL2014HPODMR, HPODMR2019CPL, HPODMR2019Berkeley, HPODMR2019CUHK, HPODMR2019CNRS}.
With synchronized control and readout sequences,  NV centers inside DAC are able to extract ac magnetic field signal \cite{PR2021HPODMR}, a prerequisite to implement NMR measurement.
Furthermore, the near-perfect spin polarization of NV electron spins offer an opportunity to build hyperpolarization of nearby nuclear spins, which can dramatically improve the NMR signal of nanoscale samples \cite{NV_DNP_NMR}.
In this letter, we demonstrate that these merits of diamond NV centers can be fully integrated into the high-pressure environment,
providing an alternate solution to the sample volume issue of high-pressure NMR.
By working near the excited state level anti-crossing (ESLAC) of diamond NV centers, $^{14}$N nuclear spin of a micro-diamond (1~femtoliter in size) are polarized, coherently manipulated, and read out through a simple sequence combining laser and radio-frequency (RF) pulses.
Hyperpolarization-enhanced NMR spectra are measured up to 16.6 GPa, and unexpected pressure shift of $^{14}$N nuclear quadrupole and hyperfine coupling terms are observed.

We start by introducing nuclear spin hyperpolarization enabled by diamond NV centers.
An NV center is formed by a substitutional nitrogen atom and an adjacent vacancy of the diamond lattice (Fig. 1 (a) inset).
It is a spin-1 system with a spin triplet as the ground state. Under green laser excitation, the NV electron spin will be polarized to the $|m_S=0\rangle$ state due to spin-dependent  transition paths. Nearby nuclear spins, including the host nitrogen (${}^{14}$N ) and carbon (${}^{13}$C) nuclear spins, bring hyperfine splitting to the NV energy levels and are usually immune to the optical excitation.
However, the optical spin polarization of NV electron spin can transfer to nearby nuclear spins under specific conditions.
As shown in Fig. 1, an external magnetic field of about 500 Gauss brings the NV system to its ESLAC, where energy difference between the $|m_S=0\rangle$ and $|m_S=-1\rangle$ states disappears, enabling flip-flop process of the coupled spin system.
As a result, a short green laser pulse will polarize the system to the $|m_S=0, m_I=+1 \rangle$ state (for ${}^{14}$N ) \cite{PRL2009DNP, PRA2009DNP}.
After tuning off the laser pulse, NV electron spin will soon settle down to its ground states where the flip-flop process is prohibited by a different zero-field splitting ($D_{GS}=$2.87~GHz for ground states, and $D_{ES}=$1.42~GHz for excited states), and NMR measurement can be implemented in the NV ground states. 

The spin-dependent fluorescence at NV ESLAC also provides an efficient method to read out the nuclear spin states optically \cite{PRA2009DNP, DNP_Readout}. The $|m_S=0, m_I=+1 \rangle$ states has the largest fluorescence counts rate, and other nuclear spin states have relative smaller counts rate.  For example, the counts rate of $|m_S=0, m_I=0 \rangle$ state is about 20\% lower than that of the  $|m_S=0, m_I=+1 \rangle$ state, thus nuclear spin state can be extracted directly from the fluorescence intensity if the electron spin is at the $|m_S=0\rangle$.  Details of the fluorescence difference of the ${}^{14}$N nuclear spin states can be found in Supplemental Material  \cite{SM}.

\begin{figure}[tb]
	\includegraphics[width=0.48\textwidth]{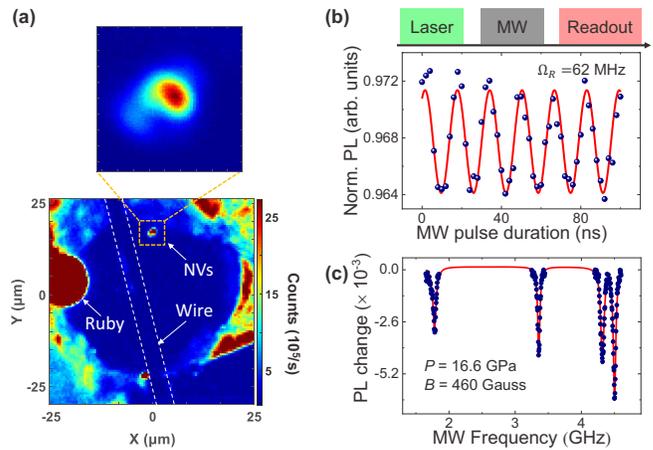}
	\caption{(color online).  Coherent control of NV electron spins and external magnetic field alignment for high-pressure NMR.  (a)  Confocal fluorescence images of the diamond anvil culet. Several microdiamonds with ensemble NV centers are loaded into the DAC culets, and a ruby particle is used for pressure calibration. A thin platinum wire (marked by the white dash line) is used to delivery microwave and radio-frequency pulses to NV centers. (b) Rabi oscillations  of NV electron spins. The measured Rabi frequency reaches 62~MHz.  (c) Optically detected magnetic resonance (ODMR) spectrum of an microdiamond as the external magnetic field is aligned to the quantization axis of the selected NV group. }
 \label{fig2}
\end{figure}

The experimental setup and sample loading methods have been described else-where \cite{HPODMR2019CPL}.
In current experiment, microdiamonds with average diameter of 1~$\mu$m are used (Adamas Nanotechnologies, NV-concentrations of about 3 ppm).
Fig. 2(a) shows confocal fluorescence images of the diamond culet after loading microdiamonds and a ruby pressure gauge.
A thin platinum wire is placed close to the microdiamonds to transmit RF and microwave (MW) pulses.
As shown in Fig. 2(b), Rabi frequency of NV electron spins reaches 62~MHz with a 16-watt microwave amplifier (Mini-circuts, ZHL-16W-43-S+), indicating the wire-based microwave transmission line works well on the diamond cutlets.

We then move to the alignment of the external magnetic field.
The magnetic field is generated by a permanent magnet, whose position and orientation are controlled by a translation stage and two rotation stages.
The strength and orientation of the magnetic field are measured by the optically detected magnetic resonance (ODMR) spectra of ensemble NV centers in the microdiamond.
There are four possible NV orientations in the diamond lattice, each of them contributes two resonances, so the full ODMR spectra exhibit eight resonances in general.
When the external magnetic field is aligned to the quantization axis of one NV group, the resonances of the other three NV groups are degenerated, resulting in an ODMR spectrum of only four dips, as shown in Fig. 2(c).
Note the zero-filed splitting ($D_{GS}$) of this spectrum has shifted to 3.116 GHz (naturally at 2.87 GHz), as a pressure of  16.6 GPa (calibrated by ruby fluoresces) is applied.
The magnetic field strength is tuned to 460 Gauss in order to polarize the ${}^{14}$N nuclear spin.
Details of the pressure and magnetic field calibrations can be found in Supplemental Material \cite{SM}.

The nuclear spin polarization cannot be observed directly on the ODMR spectra, as the widths of the ODMR resonant dips are too large.
There are several factors contribute to the line width of the ODMR spectra of ensemble NV centers.
Firstly, the power broadening of laser and microwave pulses, which can be suppressed by reducing the applied powers.
Secondly, inhomogeneous broadening of ensemble NV centers related to local strain, charge, and nuclear spin distributions \cite{ZuPRL2018}.
Thirdly, coupling to electron spin baths (mainly P1 centers), which leads short spin coherence time (about tens of nanoseconds) and is the dominate factor of the current case. The NV coherence time limited ODMR width is about several MHz, too large to resolve the hyperfine splitting of the ${}^{14}$N nuclear spin (2.16 MHz).


\begin{figure}[t]
	\includegraphics[width=0.48\textwidth]{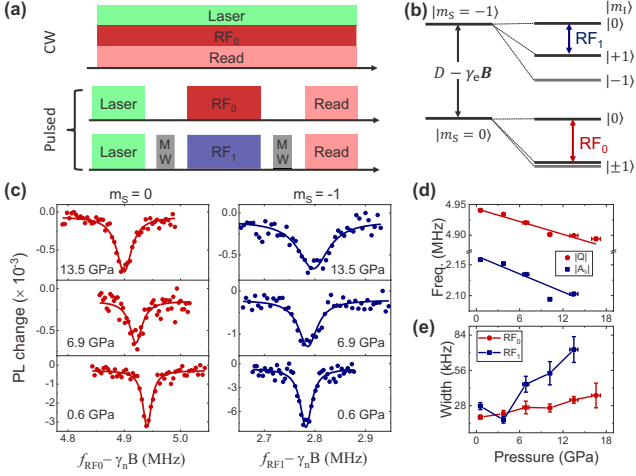}
\caption{(color online). NMR of $^{14}$N spin ensemble under high pressure.
(a) Pulse sequences to implement high-pressure NMR. The upper (lower) one shows CW (pulsed) mode.
(b) Energy levels of the coupled electron and nuclear spin system, with the transitions of the NMR measurements are labeled.
(c) NMR spectra of $^{14}$N spin ensemble under pressures of 0.6 GPa, 6.9 GPa and 13.5 GPa.
Left (right) pane: NV electron spins at the $m_s=|0>$ ($m_s=|-1>$) state and CW (pulsed) method is used.
(d) Absolute value of $Q$ (red) and $A_{\parallel}$ (blue) as function of pressure, solid lines are linear fitting, giving slopes of
$\frac{dQ}{dP}= 3.5 \pm 0.4$ kHz and $\frac{dA_{\parallel}}{dP}= 4.9 \pm 1.1$ kHz.
(e) Pressure dependence of the width of $^{14}$N NMR spectra, data in red (blue) are measured at the $m_s=|0>$ ($m_s=|-1>$) state.
}
	\label{fig3}
\end{figure}

We then demonstrate NMR, or more precisely nuclear quadrupolar resonance (NQR), of ${}^{14}$N nuclear spins under high pressures.
In the first experiment, NMR spectra are measured in a continue-wave (CW) mode.
As illustrated in Fig. 3(a), the polarization laser, driving RF pulses, and fluoresce collection are implemented simultaneously, while the RF frequency is scanned.
Under continue laser excitation, the NV electron spins are polarized and fixed to the $|m_S=0 \rangle$ state, and ${}^{14}$N nuclear spins are polarized to the $|m_I=+1 \rangle$ states if the RF pulses are off-resonance. When the RF pulse matches the resonant frequency of ${}^{14}$N nuclear spins, some population will be driven to the $|m_I=0 \rangle$ state, and a resonant dip can be observed, as shown in Fig. 3(c). This CW-NMR scheme is easy to implement but has limited contrast.
NMR spectra of this microdiamond are measured up to 16.6 GPa, and pressure-dependent shift and broadening of the NMR spectra are observed, as shown in Fig. 3(d) and (e).

An alternate and more robust approach to implement high-pressure NMR is using the pulsed mode, in which the spin polarization, manipulation and readout steps are implemented sequentially, as shown in Fig. 3(a). The RF pulse duration is chosen to achieve a $\pi$ rotation of the nuclear spin and is isolated from the optical pulses, thus the NMR contrast is enhanced as compared to that of the CW mode. Meanwhile, additional microwave pulses can be employed to control the state of NV electron, and NMR spectra at the NV $|m_S=-1 \rangle$ states can be obtained, as shown in Fig. 3(c). By fitting the NMR spectra with a single-peak Lorentz function, the resonant frequency and full width at half maximum (FWHM) are extracted, as summarized in Fig. 3(d) and (e).

The resonant frequency of the host ${}^{14}$N nuclear spins, $f_{RF0}$  and $f_{RF1}$ (as the NV electron spin set to $|m_S=0 \rangle$ or $|m_S=-1 \rangle$ state, respectively), are determined by the nuclear quadrupole coupling constant ($Q$), hyperfine coupling ($A_{\parallel}$) to the center electron spin, and the nuclear spin Zeeman effect, which can be written as:
\begin{equation*}
\begin{aligned}
f_{RF0}& = Q(P) + \gamma_n B, \\
f_{RF1}& = Q(P) + A_{\parallel}(P) + \gamma_n B,
\end{aligned}
\end{equation*}
where $\gamma_n$ is gyromagnetic ratio of ${}^{14}$N.
The nuclear Zeeman effect term is known as the strength of the external magnetic field $B$ has been measured by the ODMR spectra of NV electron spins (see Supplemental Material for details), so the combined measurement of $f_{RF0}$  and $f_{RF1}$ can be used to extracted the quadrupolar and hyperfine coupling parameters, as summarized in Fig. 3(d).

 As the pressure increases from 0.6 GPa to 16.6 GPa, the absolute value of the quadrupolar splitting decreases from 4.94 MHz to 4.89 MHz, and the hyperfine coupling term decreases from 2.16 MHz to 2.10 MHz.
 This is an unexpected result, as recent experiments have shown that thermal expansion of the diamond lattice would decrease the quadrupolar and hyperfine terms \cite{Q_A_T_PRB2020, Q_T_PRR2020}.
The quadrupolar term $Q$ is determined by the electric field gradient at the position of the nitrogen nucleus,
 and the hyperfine term $A_{\parallel}$ is determined by the Fermi contact term and dipole interaction.
 The absolute value of both terms decrease with increasing pressure, indicating the wave-function of the unpaired spin of NV centers shifts away from the nitrogen nucleus under pressures. Further experiments with nearby carbon nuclear spins, and \emph{ab initio} calculations should be pursued to fully understand the compression and modification of the NV electronic orbitals under high pressures.

As summarised in the lower pane of Fig. 3(d), the widths of the NMR spectra reaches tens of kilohertz.
We attribute the relative large NMR width to the short dephasing time of the nuclear spin ensembles.
To illustrate this point, we measure the Rabi oscillation and free-induction deday (FID) signal of the ensemble nuclear spins.
Figure 4(a) presents nuclear spin Rabi oscillation driving by RF pulses of different powers, envelop decays in a time scale of about 100~$\mu$s are observed for all the measurements, indicating a short nuclear spin coherence time of this sample.
Figure 4(b) presents nuclear spin FID signal, giving a nuclear spin dephasing time ($T_{2N}^*$) of 70 $\pm~ 10~\mu$s,
such a short dephasing time is consistent with the NMR width of about 30~kHz.

\begin{figure}[t]
	\includegraphics[width=0.48\textwidth]{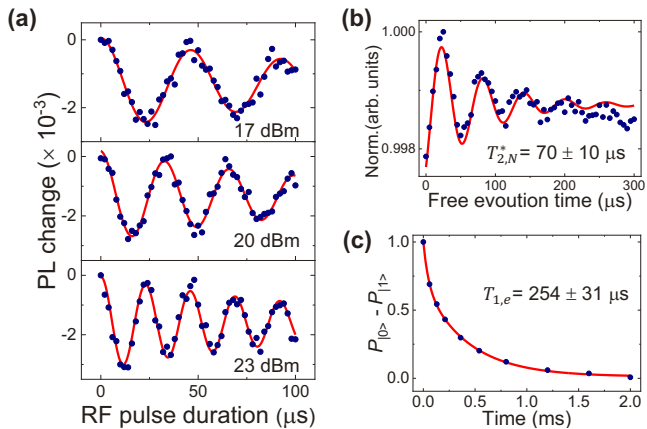}
	\caption{(color online).
Coherent control and dephasing of $^{14}$N spin ensemble under high pressure. The applied pressure is 0.6 GPa.
(a) Rabi oscillation of $^{14}$N spins at different RF drinving powers, while the electron spin is at the $m_S=0$ state. The used pulse sequence is shown in Fig. 3 (a).
(b) Free induction decay (FID) of $^{14}$N spins. The red line is fitting of a cosine function with exponential decay, which gives nuclear spin dephasing time ($T_{2N}^*$).
(c) Spin relaxation time of NV electron, which sets the coherence time of its nearby nuclear spins. The red line is fitting of an exponential decay.
}
	\label{fig4}
\end{figure}

It has been shown that the dephasing time of a nuclear spin is constrained by a fast relaxation nearby electron spin, as a random flip of the central spin brings a random phase to the strongly coupled nuclear spin \cite{C13_T2_1s}. To verify whether this mechanism dominate the decoherence of the measured nuclear spins, we further measure the NV electrons spin relaxation time, as presented in Fig. 4(c). An exponential decay function fitting of this signal gives $T_{1e}$ of 254 $\pm ~31~\mu$s, comparable to the nuclear spin dephasing time.
We note the measured $T_{1e}$ is much shorter than the results of our previous experiments \cite{HPODMR2019CPL}, even at a relative low pressures of about 0.6 GPa.
The unusually short NV spin relaxation time is partially caused by the dense electron spin bath (P1 centers) in this microdiamond.
Besides, there is an unknown and pressures dependent factor, as the NMR width increases with pressure (see Fig. 3 (d)).
Along with the broadening effect, the contrast of NMR spectra decreases with pressures,
and only 0.01\% contrast is left at pressure of 16.6 GPa.

It is expected that using diamond NV center with better spin coherence could improve the spectrum resolution of high-pressure NMR, and higher working pressures are achievable.
We would also like to point out that the restrictions on the width of high-pressure NMR by NV probes itself can be fully overcame with recently developed correlation measurements.
By synchronizing the applied sequence of pulses to a stable oscillator, the width of the NMR spectra is no longer limited by the relaxation time the NV electron spin, and spectrum resolution of better than 1~mHz has been demonstrated in experiments \cite{Corr_2017_degen, Corr_2017_Fedor}.
Furthermore, considering the quantum nature of the target nuclear spins, sequential weak measurements can be used to overcome the impact of back action inherent of quantum measurement \cite{Corr_weak_WP, Corr_weak_degen}.

 Note that the NV-based dynamic nuclear polarization is a fast developing field, and there are other efficient protocols for transferring the spin polarization of NV center to nearby nuclear spins, including Hartmann-Hahn resonant driving \cite{GQ2014NVDNP, NVDNPPRL}, PulsePol method \cite{PulsePol}, hyperpolarization at ground state level anti-crossing \cite{DNP_LAC}, and so forth. These dynamic nuclear polarization protocols are all compatible with the DAC high-pressure environment. However, the above-mentioned direct fluorescence readout method is not applicable to most of the dynamic nuclear polarization protocols. Nevertheless, after building the hyperpolarization of nuclear spins, nearby NV centers can work as in-situ sensitive probes to detect the nuclear spin procession signal inside the DAC \cite{PR2021HPODMR}.

 The combination of NV-based dynamic nuclear spin polarization and in-situ spin detection provides a perfect solution to the sample volume issue of high-pressure NMR.
 In current experiments, the sample volume is only about 1 $\mu m^3$ (1 femtoliter) and can be further deceased.
 Such a small sample volume is compatible with the small culets ($\leq$ 100 $\mu$m) of megabar region DAC.
 Meanwhile, a small detection volume would effectively eliminate inhomogeneous broadening due to gradients of pressure, magnetic field, and RF driving fields.
 Altogether, the NV-based scheme would enable megabar region NMR with micrometer spatial resolution, and can be
 a powerful tool to study exotic states of condensed matter physics like superconductivity of hydrides under high pressures \cite{LaH10_HP_PRL, LaH10_HP_Nature, H_NMR200GPa}.
 To fully exploit this approach, a detailed study of the optical and spin properties of diamond NV center under ultrahigh pressure is in progress.

In summary,  we propose and demonstrate a high-pressure NMR (and NQR) scheme enabled by diamond NV centers.
The first critical step of this scheme is hyperpolarization of the target nuclear spins with NV centers, followed by standard NMR sequence to control the evolution of the nuclear spins, and a second critical step is detection of the nuclear spin states with nearby NV quantum sensors.
We experimentally demonstrate high-pressure NMR of ${}^{14}$N nuclear spins in a microdiamond up to 16.6 GPa.
NMR spectra reveal that both the ${}^{14}$N nuclear quadrupole and hyperfine coupling terms deceases with increasing pressures, indicating the electron probability density of NV centers shifts away from the nitrogen nucleus under pressures.
The tens of kilohertz width of the NMR spectra is attributed to the fast relaxation time of the NV electron spins and can be suppressed with high-purity diamond and advanced correlation method.
These results deepen our understanding of the electronic structure of diamond NV center, and shine a new light on NMR and NQR under extreme conditions.

\emph{Acknowledgements }
This work was supported by Beijing Natural Science Foundation (Grant No. Z200009), Chinese Academy of Sciences (Grant Nos. YJKYYQ20190082, XDB28000000, XDB33000000, XDB25000000, QYZDBSSW-SLH013), the National Natural Science Foundation of China (Grant Nos. 11974020, 12022509, 12074422, 11934018, T2121001), the National Key Research and Development Program of China (Grant Nos. 2019YFA0308100, 2021YFA1400300, 2018YFA0305700), the Youth Innovation Promotion Association of Chinese Academy of Sciences (Grant No. 202003). This work was partially carried out at high-pressure synergetic measurement station of Synergic Extreme Condition User Facility.


%

\end{document}